\documentclass[a4paper,12pt]{revtex4}
\usepackage[dvips]{graphicx}
\usepackage{amsmath}
\usepackage{amssymb}
\usepackage{url}
\usepackage{color}
\usepackage{esint}
\usepackage{float}
\usepackage{calc}
\usepackage{ae}
\usepackage{epstopdf}
\usepackage{epsfig}

\newcommand{\be}{\begin{equation}}
\newcommand{\ee}{\end{equation}}
\newcommand{\bea}{\begin{eqnarray}}
\newcommand{\eea}{\end{eqnarray}}

\newcommand{\p}{\partial}

\newcommand{\etal}{{\it et al}}

\newcommand{\rd}{\ensuremath{\mathrm{d}}}
\newcommand{\sub}[1]{\ensuremath{_{\text{#1}}}}
\newcommand{\Ordo}{\ensuremath{\mathcal{O}}}

\begin{document}


\begin{center}
{\bf \Large {\sffamily Numerical calculation of ion runaway distributions}}
\vspace{0.25cm}\\
{\sffamily O. Embr\'eus$^1$, S. Newton$^2$, A. Stahl$^1$, E. Hirvijoki$^1$, T. F\"{u}l\"{o}p$^1$}\\
{\footnotesize {\it $^1$ Department of Applied Physics, Chalmers University of Technology, SE-412 96 G\"{o}teborg, Sweden}} \\
{\footnotesize {\it $^2$ CCFE Fusion Association, Culham Science Centre, Abingdon, Oxon, OX14 3DB, United Kingdom}}\vspace{0.15cm}
\end{center}

\begin{abstract}
  Ions accelerated by electric fields (so-called runaway ions) in
  plasmas may explain observations in solar flares and fusion
  experiments, however limitations of previous analytic work have prevented
  definite conclusions.  In this work we describe a numerical solver
  of the 2D non-relativistic linearized Fokker-Planck equation for
  ions. It solves the initial value problem in velocity space with a
  spectral-Eulerian discretization scheme, allowing arbitrary plasma
  composition and time-varying electric fields and background plasma
  parameters.  The numerical ion distribution function is then used to
  consider the conditions for runaway ion acceleration in solar flares
  and tokamak plasmas.  Typical time scales and electric fields
  required for ion acceleration are determined for various plasma
  compositions, ion species and temperatures, and the potential for
  excitation of toroidal Alfv\'en eigenmodes  during tokamak disruptions
  is considered.
\end{abstract}

\maketitle


\section{Introduction}
\label{secintro}

The phenomenon of particle runaway in a plasma is well known,
occurring in both space and laboratory
plasmas~\cite{holman1995,hollmannetal2013}.  It arises because the
friction force experienced by a charged particle decreases with
particle energy, so that the presence of a sufficiently strong induced electric
field can allow the particle to be accelerated - or {\it runaway} - to
high energy.

Electron runaway~\cite{kulsrudetal1973} has been extensively studied
in the context of magnetic confinement fusion in tokamaks, where it
can lead to the formation of localized high-energy beams which must
be carefully controlled~\cite{lehnenetal2011}.
The standard analytic method used to determine the initial evolution
of the electron distribution function in a fully ionized plasma was
introduced by Kruskal and Bernstein~\cite{kruskalbernstein1962} and
later generalized by Connor and Hastie \cite{connorhastie1975}. It
takes the form of an asymptotic expansion of the electron kinetic
equation in the electric field strength.
Once a fast electron population -- known as the primary
distribution -- is established, it can rapidly produce further fast electrons through
large-angle, or {\it knock-on}, collisions~\cite{rosenbluthputvinskii1997}. This avalanche process
leads to the so-called secondary runaway-electron generation, which is often dominant.

Ion runaway has long been of interest in the astrophysical community,
where it is thought to contribute to the observed abundance of high
energy ions in solar flares~\cite{holman1995}. It has also been used
to study the behavior of lightning channels~\cite{fuloplandreman2013}
and was observed in laboratory plasmas, e.g. in the Mega Ampere
Spherical Tokamak (MAST)~\cite{helanderetal2002} and in the Madison
Symmetric Torus~\cite{eilerman}.
The detailed mechanism of ion runaway differs from that of electron
runaway.  Friction with the electrons, which are also drifting in the
electric field, acts to cancel a portion of the accelerating field.
In the ion rest frame, the electrons have net motion anti-parallel to
the electric field, and a test ion will experience two main forces 
beyond friction against the background of charged particles:
acceleration in the electric field, and friction due to the electron drift. These
forces scale differently with ion charge, and the dominant force
is either electron friction -- with the consequence that the ions are dragged along with
the electrons -- or acceleration by the electric
field. In a pure plasma the cancellation of electric field
acceleration and electron friction is complete, but the presence of impurities, 
neutrals or effects such as particle trapping in a
non-uniform confining magnetic field allow a finite effective field
to remain~\cite{gurevich1961,furthrutherford1972}. 

Furth and Rutherford~\cite{furthrutherford1972} generalized earlier
treatments by adopting a similar expansion procedure to that used to
study electron runaway.  They determined the steady state ion
distribution function in conditions typical of operational fusion
plasmas.  Helander~\etal.~\cite{helanderetal2002} then considered the
initial value problem resulting from the onset of an accelerating
electric field, produced for example by a plasma instability. An
analytic solution for the initial time evolution of the accelerated ion
distribution function was developed, but it was noted that its
application was limited due to the low electric fields required for
it to be valid.  Both of these ion runaway studies considered only the
presence of a trace impurity population, consistent with typical
operating conditions in fusion plasmas.

Plasmas with significant impurity content are also common, however.
Astrophysical plasmas often consist of a mixture of dominant species,
as well as containing trace
elements~\cite{holman1995,schmelzetal2012}.
Disruptions~\cite{hollmannetal2013}, which are a common cause of
electron runaway in tokamaks, are also typically associated with an
increase in impurity content, either due to deliberate gas injection
for mitigation purposes or due to plasma-wall interaction. Therefore,
in Ref.~\cite{fulopnewton2014}, the initial value formulation of the
problem posed in Ref.~\cite{helanderetal2002} was extended to account for arbitrary
plasma composition. The potential for ions accelerated during a
disruption to excite low frequency plasma instabilities was
considered analytically. The results were inconclusive since the
asymptotic expansion used to develop the analytic solution was not
strictly valid for disruption-type parameters.  The limitations of the
analytic solutions available in previous work motivate the
development of a numerical tool to allow detailed study of the time
evolution of an ion runaway distribution.

Here we describe the formulation and implementation of an efficient
finite-difference--spectral-method tool, \textsc{CODION} (COllisional
Distribution of IONs), that solves the two-dimensional momentum space
ion kinetic equation in a homogeneous plasma.  \textsc{CODION} solves
the ion Fokker-Planck equation as an initial value problem and allows
for time-variation of the electric field and bulk distribution
parameters (temperature, density, charge and mass) of each plasma
species independently.  Due to its speed it is highly suitable for
coupling within more expensive calculations, e.g. studies of
instabilities driven by the fast ions or comprehensive modeling of ion
acceleration with self-consistent coupling to solvers of Maxwell's
equations. Using \textsc{CODION} we obtain illustrative
two-dimensional ion velocity space distributions, which demonstrate
the typical behaviour of runaway ions in a variety of physical
scenarios. We show that during typical tokamak disruptions ions are
unlikely to be accelerated to velocities high enough for resonant
interaction with toroidal Alfv\'en eigenmodes (TAE). Therefore the
experimentally observed TAE activity cannot be explained by the ion
runaway mechanism.

The rest of the paper is organized as follows.  In Section
\ref{secridistrib} we describe the ion Fokker-Planck
equation, and in Section \ref{secCODION} we outline its numerical
implementation in \textsc{CODION}. In Section \ref{seccolops} we
explore the numerical solution, including the effect of various
approximations to the collision operator.  In Section
\ref{applications} the model is applied to a variety of physical
scenarios, illustrating typical acceleration time scales in laboratory
and space plasmas.
Finally we close with concluding remarks in Section \ref{conclusions}.


\section{Runaway Ion Distribution}
\label{secridistrib}
We consider the problem of ion acceleration by induced electric fields
with a component parallel to the background magnetic field in a
plasma.  We restrict ourselves to straight field line geometry, and
assume a homogeneous background plasma.  The time evolution of the ion
distribution is then given by the Fokker-Planck equation, and particle
acceleration will be opposed by various friction forces.
The effect of friction with neutral particles can be significant in
various physical situations, for example giving rise to charge
exchange losses, which was studied in the context of lightning
discharges in Ref.~\cite{fuloplandreman2013}. Here we will focus
on fully ionized quasi-neutral plasmas, in which case the
friction is the result of inter-species Coulomb collisions only.

We are interested in the initial value problem where an electric field
appears in what was previously an equilibrium state.  Therefore we assume that
the initial particle distribution functions are stationary Maxwellians
$f_{M}$ -- possibly at different temperatures -- and consider their distortion from this state by the electric
field.  We linearize the collision operator about this background Maxwellian, and
neglect the non-linear contribution to the evolution.  This restricts
the study to situations where only a small fraction of the ion
population is accelerated, or to the initial stages of ion runaway.  A
reason for why this is sufficient in the present context is that, once
a high energy population forms, the runaway ions have the potential to
excite instabilities~\cite{fulopnewton2014}, which will have a strong
impact on the further evolution of the distribution.  Note that the
non-linear terms of the kinetic equation are sometimes required to account for
the transfer of energy from the electric field into heating the
distribution.  These aspects of the longer term evolution of the
distribution are beyond the scope of the work presented here.

The linearized collision operator consists of the test and
field-particle terms.  Momentum and energy conservation in ion
self-collisions is retained by approximating the field-particle piece
with restoring terms, as described in Ref.~\cite{abeletal2008}, without 
the need to include the full field-particle operator which is numerically more expensive.  Collisions
with the other ion species could be treated similarly, however this
would require the simultaneous evolution of the distribution functions
of multiple species.  Therefore, we consider only the evolution of the
ion species with the highest runaway rate, so that the other ion
species remain approximately stationary and only the test-particle
piece of the unlike ion collision operator needs to be retained. While
it is difficult to verify \emph{a priori} that this condition is
satisfied, it can be determined by numerical simulation of each ion
species individually, assuming the others to remain stationary. Because
of the sensitivity of acceleration rate to ion charge and mass (as demonstrated in Section \ref{applications}), the condition can typically
be well satisfied as the acceleration rate of different species is often separated by several orders of magnitude. 

The velocity-dependent friction on a test particle resulting from
collisions with a Maxwellian background species has a peak near the
thermal velocity of the background due to the form of the Coulomb
interaction between charged particles.  In the case of electrons, the
friction force will be a monotonically decreasing function for
velocities above the electron thermal velocity, allowing an electron
to run away to large energy (where
relativistic~\cite{connorhastie1975} and synchrotron
effects~\cite{adamprl} become dominant).
We focus on situations where the ion, $i$, and electron, $e$, thermal
velocities satisfy $v_{Te} \gg v_{Ti}$, meaning that their temperatures are
sufficiently similar that $T_e / T_i \gg \sqrt{m_e / m_i}$.  Then, if
an ion is accelerated away from the bulk, friction against the
electron population will increase with velocity. This has the
consequence that the ion acceleration will be naturally balanced by
the electron friction for some $v<v_{Te}$, if the electric field is
below a threshold value similar to the Dreicer field
\cite{dreicer1959} for electrons.

Since an electron reacts to the electric field on a time-scale
$m_e/m_i$ times shorter than the ions, we assume the electron
population to be in a quasi-steady state on all time-scales of
interest for ion acceleration. Parallel force balance for the electron
distribution then requires that the total electron-ion friction
cancels the acceleration by the electric field, $n_e eE_\parallel =
\sum_j {R}_{ej\parallel} = \sum_j \int \rd \bold{v}\, m_e v_\parallel
C_{ej}$, where the sum is taken over all ion species $j$ in the
plasma.  Due to the small mass ratio, the electron-ion interaction is
dominated by pitch-angle scattering, so that $C_{ej} = C_{ei} n_j
Z_j^2/n_i Z_i^2$, and we can solve for the friction between electrons
and the ion species of interest, $R_{ei\parallel} = (n_i
Z_i^2/Z\sub{eff})eE_\parallel$, where the effective charge is
$Z\sub{eff} = \sum_j n_j Z_j^2/n_e$.

The electron distribution can be written as $f_e = f_{Me}+f_{e1}$,
with $f_{Me}$ a Maxwellian distribution drifting with the bulk ion
velocity ${\bf V}_i$, and a correction $f_{e1}$ varying over
velocities of order $v_{Te}$, accounting for the electron drift
behavior in the electric field. Then the linearized ion-electron
collision operator, neglecting terms quadratic in $f_{e1}$, can be
simplified~\cite{perbook,holman1995}, noting that momentum
conservation in binary collisions requires that ${\bf R}_{ie} = - {\bf
  R}_{ei}$, \be C_{ie}\left\{f_i,f_e\right\} = \frac{{\bf R}_{ei}}{m_i
  n_i} \cdot \frac{\p f_i}{\p {\bf v}} +
C_{ie}\left\{f_i,f_{Me}\left({\bf v} - {\bf V}_i\right)\right\}.
\label{eqcei}
\ee
The first term, which describes the friction arising
from the drifting electron distribution and was calculated above, readily combines with that
describing acceleration by an electric field, giving rise to the
effective electric field $E^* = E_\parallel - R_{ei\parallel}/n_i Z_i
e = (1-Z_i/Z\sub{eff})E_\parallel$.  Thus, as noted in the
Introduction, in a pure plasma where $Z_i = Z_{\rm eff}$, net ion
acceleration will not occur.  Light ions with $Z_i < Z_{\rm eff}$ can
be accelerated in the direction of the electric field.  Heavy
impurities with $Z_i > Z\sub{eff}$ will be accelerated in the opposite
direction, that is in the direction of electron runaway (the latter
case was studied by Gurevich~\cite{gurevich1961}).  The second term in
Eq.~(\ref{eqcei}) describes the slowing down of the fast ions on the
electrons, as well as the slow energy exchange between the bulk
species. Note that the ion flow velocity correction is time dependent
and formally small in the runaway density. This term will become more
significant in the ion-electron momentum exchange as the runaway
distribution builds up.

Finally then, the kinetic equation we consider for the evolution of
the ion distribution function in the presence of an accelerating
electric field and arbitrary plasma composition is the following:
\be
\frac{\p f_i}{\p t} + \frac{Z_ie}{m_i} E^* \left(\xi\frac{\p}{\p v} + \frac{1-\xi^2}{v}\frac{\p}{\p \xi}\right)f_i = \sum_s C_{is}\{f_i\},
\label{eqgenkeforevolution}
\ee
where $\xi= v_\parallel / v$, and the effect of collisions with the
background Maxwellian populations is described by the sum over all
particle species $s$ in the plasma,
\bea
\sum_s C_{is}\{f_i\} &=& \frac{1}{\tau_{ie}}\sum_s\frac{n_s Z_s^2}{n_e}\left\{\frac{\phi\left(x_s\right) - G\left(x_s\right)}{2x_i^3}\frac{\p}{\p \xi}\left[\left( 1 - \xi^2\right) \frac{\p f_i}{\p \xi}\right]\right.\nonumber \\ & & \left.\hspace{1.2cm}+ \frac{1}{x_i^2}\frac{\p}{\p x_i} \left[2 \frac{T_i}{T_s}x_i^2G\left(x_s\right) f_i + x_i G\left(x_s\right)\frac{\p f_i}{\p x_i}\right]\right\} \nonumber \\
& & + \frac{1}{\tau_{ii}}\left[2\nu_s(v) x_i \xi u_\parallel + \nu_E(v) x_i^2 Q\right]f_{Mi},
\label{eqioncolop}
\eea
where $\tau_{is}^{-1} = n_s e^4 Z_i^2 Z_s^2 \ln \Lambda / 4\pi
\epsilon_0^2 m_i^2 v_{Ti}^3$, $x_s = v / v_{Ts} = \sqrt{m_s v^2/2T_s}$
and the usual error function $\phi(x) = (2/\sqrt{\pi})\int_0^x \rd y\,
e^{-y^2} $ and Chandrasekhar function $G(x) = \left[\phi(x) - x
  \phi^\prime(x)\right]/2x^2$ appear.  The dimensionless moments
$u_\parallel$ and $Q$ of the distribution function appearing in the
momentum and energy restoring terms of the self-collision operator are
\be
u_\parallel\{f_i\} = \frac{3}{2}v_{Ti}\frac{\int \rd^3v \,\nu_s(v) v_\parallel f_i}{\int  \rd^3v \,\nu_s(v) v^2 f_{Mi}}, \hspace{0.7cm}
Q\{f_i\} = v_{Ti}^2\frac{\int  \rd^3 v \,\nu_E(v) v^2 f_i}{\int  \rd^3v \,\nu_E(v) v^4 f_{Mi}},
\ee
with the scattering frequencies given by
\be
\nu_s(v) = 4 \frac{G(x_i)}{x_i}, \hspace{0.7cm}
\nu_E(v) = 2\left(4 \frac{G(x_i)}{x_i} - \frac{\phi(x_i)}{x_i^3}\right).
\ee

\begin{figure}[htbp]
\begin{center}
\includegraphics[width=0.49\textwidth]{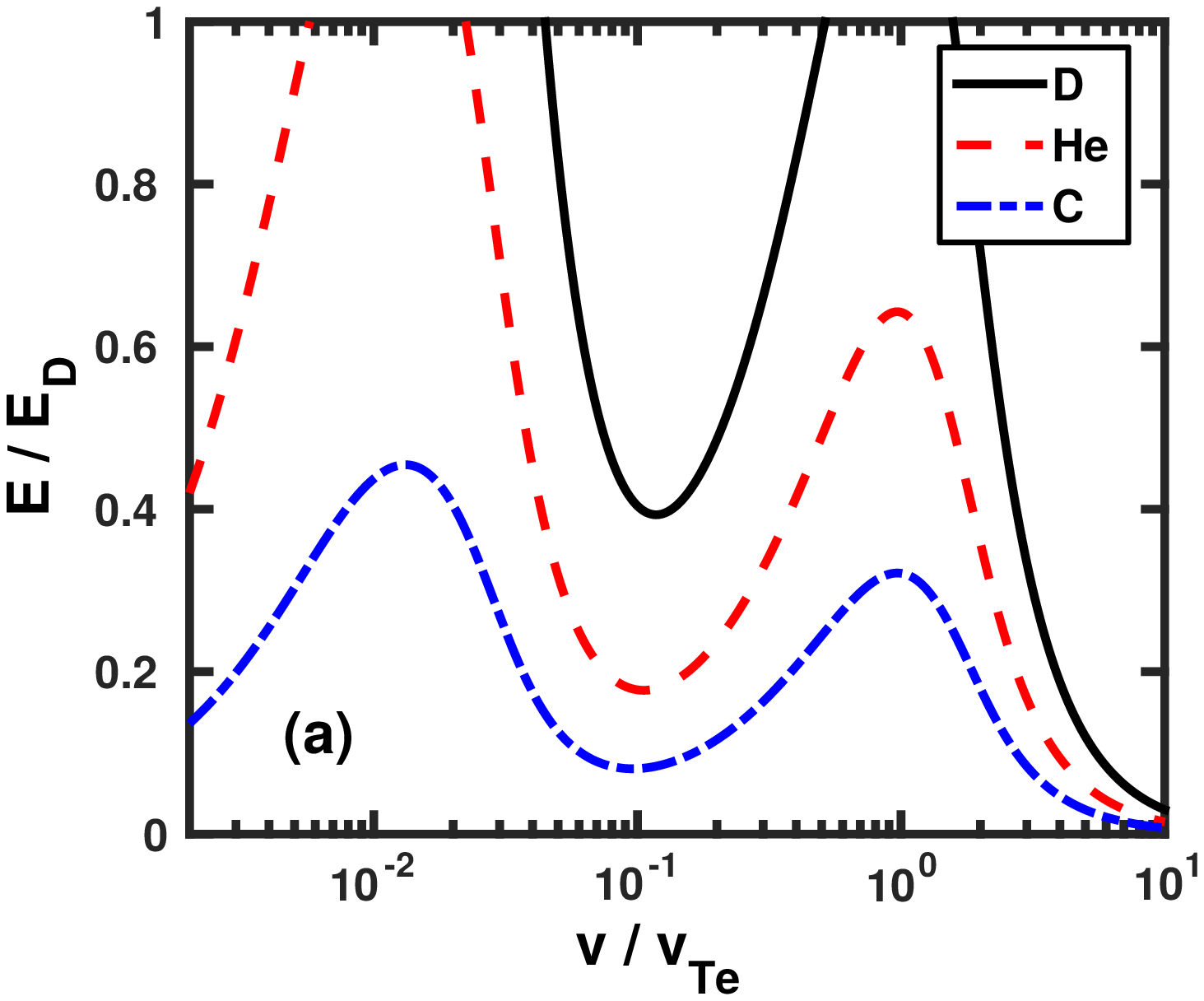}
\includegraphics[width=0.49\textwidth]{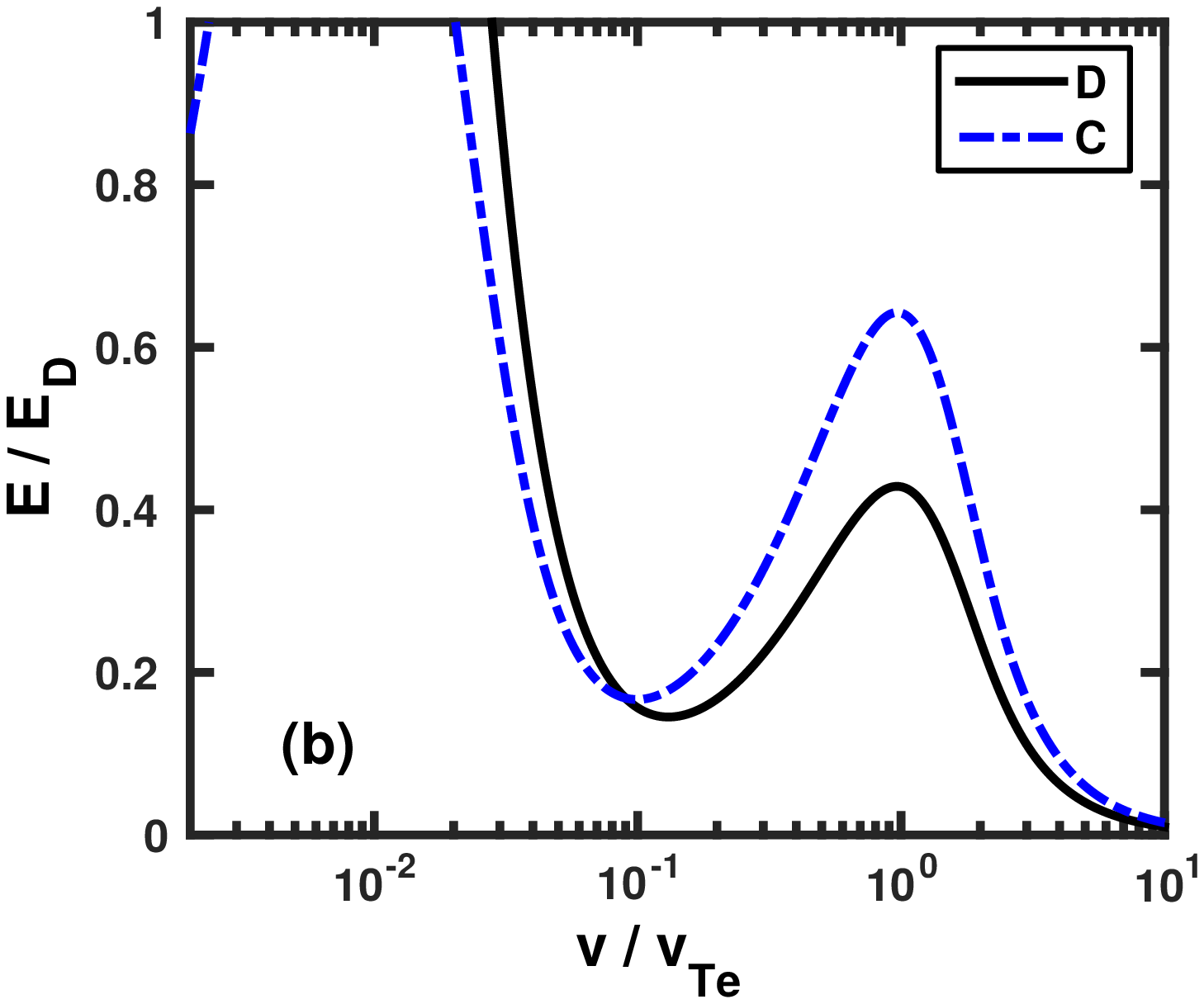}
\caption{Electric field needed to accelerate fully ionized test
  particles of various ion species in an impure deuterium plasma, as a
  function of velocity. (a) $n\sub{C}/n\sub{D} = 0.4\%$,
  $n\sub{He}/n\sub{D} = 5\%$, $Z\sub{eff} = 1.2$ and (b)
  $n\sub{C}/n\sub{D} = 4\%$, $n\sub{He}/n\sub{D} = 5\%$, $Z\sub{eff} =
  2$, and all particle species are taken to be at the same
  temperature. Since $Z\sub{eff} = Z\sub{He}$ in (b), no electric
  field for which the model is valid will accelerate helium ions. The
  quantities shown are independent of the density and temperature of
  the plasma.}
\label{figtestedemo}
\end{center}
\end{figure}

The runaway behavior of interest can be demonstrated by considering
the simpler dynamics of an ion test
particle~\cite{fuloplandreman2013,hintonreview} in the presence of the
electric field. The ion equation of motion takes the form $m_i \rd
v_\parallel/\rd t = Z_ieE^* + F^\text{test}_i$, where the test
particle collisional friction is given by
\be
F^\text{test}_i(v) = - Z_i^2 e E_D  \sum_s \frac{n_s Z_s^2}{n_e} \frac{T_e}{T_s}\left(1 + \frac{m_s}{m_i}\right)G(x_s),
\label{eqtestfriction}
\ee
and $E_D = n_e e^3 \ln \Lambda / 4 \pi \epsilon_0^2 T_e$ is the
Dreicer field.  Thus ion acceleration can occur when $E/E_D >
F^\text{test}_i/Z_ieE_D\left|1-Z_i/Z_{\rm eff}\right|$. The required
electric field values are illustrated in Fig.~\ref{figtestedemo}, for
low and high effective charge. The figures illustrate the
non-monotonic ion friction force, with one maximum near the thermal
velocities of the ion species, and another near the electron thermal
velocity.
Therefore, as first described by Furth and
Rutherford~\cite{furthrutherford1972}, for a sufficiently strong
electric field we may expect ions to be accelerated from their initial
velocity to a higher velocity at which friction on the electrons
dominates, giving rise to a suprathermal population in the plasma.
We will compare the characteristics of the test-particle behavior,
governed by the friction force illustrated in Fig.~\ref{figtestedemo},
to the numerical solution of Eq.~(\ref{eqgenkeforevolution}) in the
next section.

Figure~\ref{figtestedemo} also illustrates that the electric fields
needed to accelerate ions are highly sensitive to ion charge and
plasma composition, due to their effect on the effective electric
field ${E^* = (1-Z_i/Z\sub{eff})E}$.
Note that the electric fields needed to accelerate ions beyond the
electron thermal velocity are significantly larger than the minimum
electric field necessary for acceleration.  Such strong fields will
not be considered here, since the validity of the linearization
typically breaks down before the ions reach a significant fraction of
the electron thermal velocity.


\section{CODION}
\label{secCODION}
In this section, we outline the implementation of
Eq.~(\ref{eqgenkeforevolution}) in the numerical tool \textsc{CODION},
which solves the ion Fokker-Planck equation numerically as an initial
value problem to give the evolution of the ion distribution function
in the presence of an accelerating electric field. It uses a
continuum-spectral discretization scheme based on that used in CODE
\cite{codepaper}. Illustrative solutions for a tokamak-like plasma are
presented, and a comparison of the obtained distribution function is
made with the behavior predicted by test-particle equations,
demonstrating the importance of collisional diffusion for the runaway
of ions. In addition we investigate the effect of different choices
for the self-collision field-particle operator.

The pitch-angle dependence of the distribution function is represented
by a truncated Legendre polynomial expansion, while velocity is
discretized on a uniform grid $v=v_n = n \Delta v$,
$n=0,\,1,\,...,\,N-1$:
\begin{align}
  f_i(v_n,\,\xi,\,t) = \sum_{l=0}^{l\sub{max}} f_l(v_n,\,t) P_l(\xi),
\end{align}
where the Legendre polynomials $P_l$ obey the orthogonality relation
$\int_{-1}^1 \rd \xi\, P_l(\xi) P_{l'}(\xi) =
{\delta_{l,l'}2/(2l+1)}$, and
\begin{align}
f_l(v_n,\,t) = \frac{2l+1}{2}\int_{-1}^1 \rd \xi \, P_l(\xi) f_i(v_n,\,\xi,\,t).
\end{align}
The integral operation $(L+1/2)\int_{-1}^1 \rd \xi \, P_L(\xi) ...$ is
applied to the kinetic equation in Eq.~(\ref{eqgenkeforevolution}) for
each $L$, producing a linear set of equations for the quantities
$f_L(v,\,t)$, using the boundary condition $f_{l\sub{max}}(v,\,t) = 0$
for all $v$. Well-known recurrence relations for the Legendre
polynomials are used to obtain analytic expressions for all the terms
appearing in the equation. For example, the Legendre polynomials are
eigenfunctions of the linearized collision operator, while the
electric field-term will produce a coupling between $f_L$ and
$f_{L\pm1}$ modes. This procedure exactly captures number conservation
for any choice of $l\sub{max} > 1$. The velocity derivatives appearing
in the kinetic equation are represented with five-point stencils
\begin{align}
\frac{\partial f_l}{\partial v}\Big|_{v_n} &= \frac{1}{12\Delta v}\sum_{m=0}^{N-1} \left( -\delta_{n,m-2} + 8 \delta_{n,m-1} - 8 \delta_{n,m+1} + \delta_{n,m+2} \right)f_l(v_m), \\
\frac{\partial^2 f_l}{\partial v^2}\Big|_{v_n} &= \frac{1}{12\Delta v^2}\sum_{m=0}^{N-1} \left( -\delta_{n,m-2} + 16 \delta_{n,m-1} - 30\delta_{n,m}+16\delta_{n,m+1}-\delta_{n,m+2}\right) f_l(v_m),
\end{align}
formally introducing an error of order $\Ordo(\Delta v^4)$. The
integral moments of the ion distribution appearing in the
self-collision model operator are discretized with a quadrature of the
form $\int \rd v \, A(v) = \sum_m w_m A(v_m)$, where the quadrature
weights $w_m$ are chosen according to Simpson's rule, also with error of
order $\Ordo(\Delta v^4)$. 

For the distribution function to be single-valued and smooth at the
origin, we enforce the boundary condition $f_l(0) = 0$ for all $l >0$
and $\rd f_0/\rd v|_{v=0} = 0$.  Since we restrict ourselves to cases
where electron friction will dominate the electric field at
sufficiently high velocities, the maximum resolved velocity can always
be chosen so that only insignificant numbers of particles are near the
edge of the grid, minimizing the effect of the choice of boundary
condition. We use the Dirichlet boundary condition $f_l(v_N) = 0$
for all $l$ at the maximum velocity. A detailed investigation of the convergence properties of the
solutions with respect to discretization parameters is described in
Ref.~\cite{olathesis}. 

The discretization procedure outlined above casts the kinetic
equation, Eq.~(\ref{eqgenkeforevolution}), into the form
\begin{align}
  \frac{\partial F}{\partial t} + MF = 0,
\end{align}
where $M$ is a sparse matrix and $F$ is a vector representing
the discretized distribution function.  Time integration is performed with the
first order implicit scheme
\begin{align}
F(t_{n+1}) = \left[I + \Delta t M(t_{n+1})\right]^{-1}F(t_n),
\end{align}
where any time-dependence of the operator $M$ is due to 
time-variation of electric fields and background plasma parameters. An
arbitrary plasma composition is determined by a set of input vectors
containing particle masses $m_s$, corresponding charge states $Z_s$,
charge densities $\rho_s = n_s Z_s$ and temperatures $T_s$.

\begin{figure}[b]
\begin{center}
\includegraphics[width=0.463\textwidth]{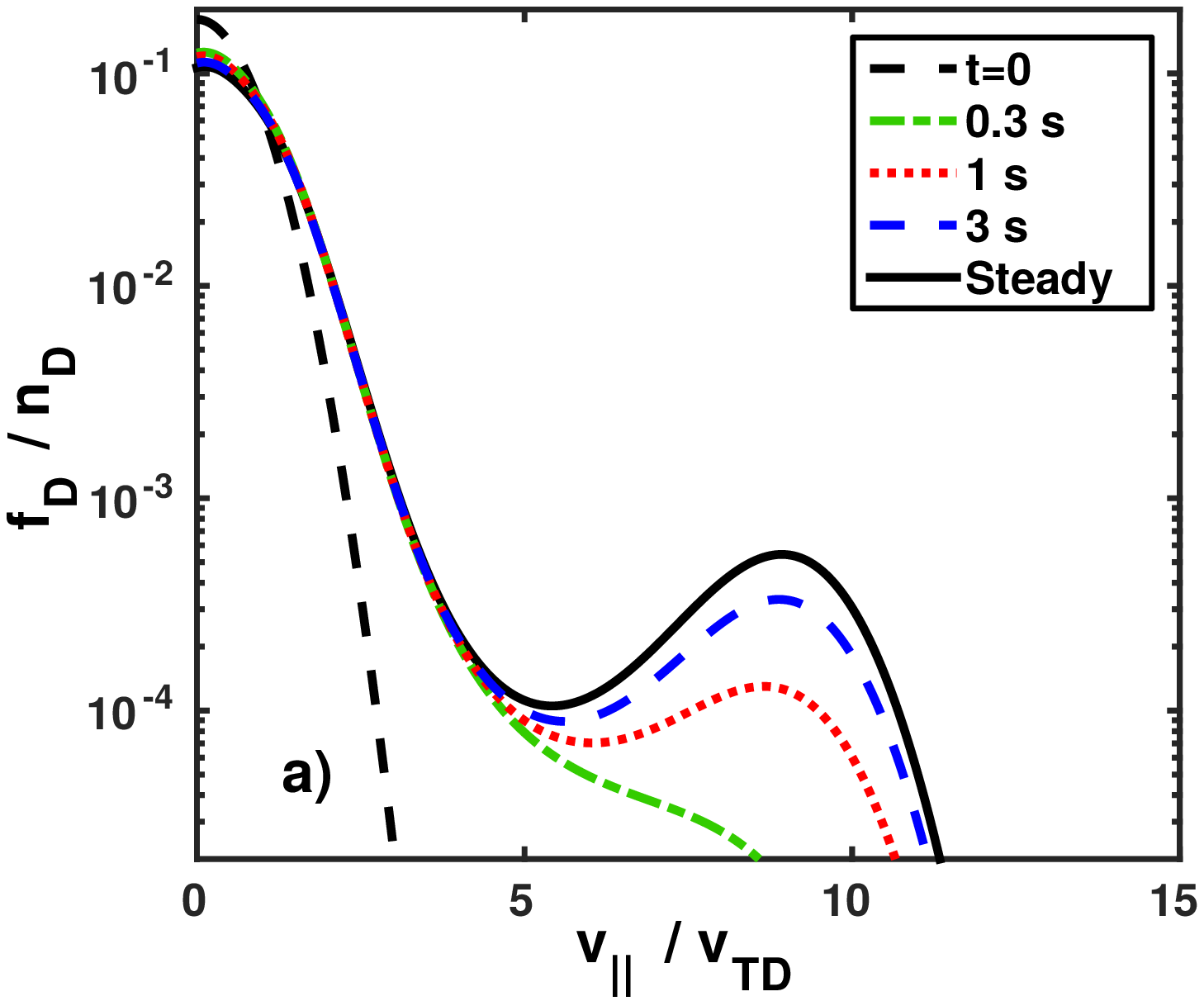}
\includegraphics[width=0.53\textwidth]{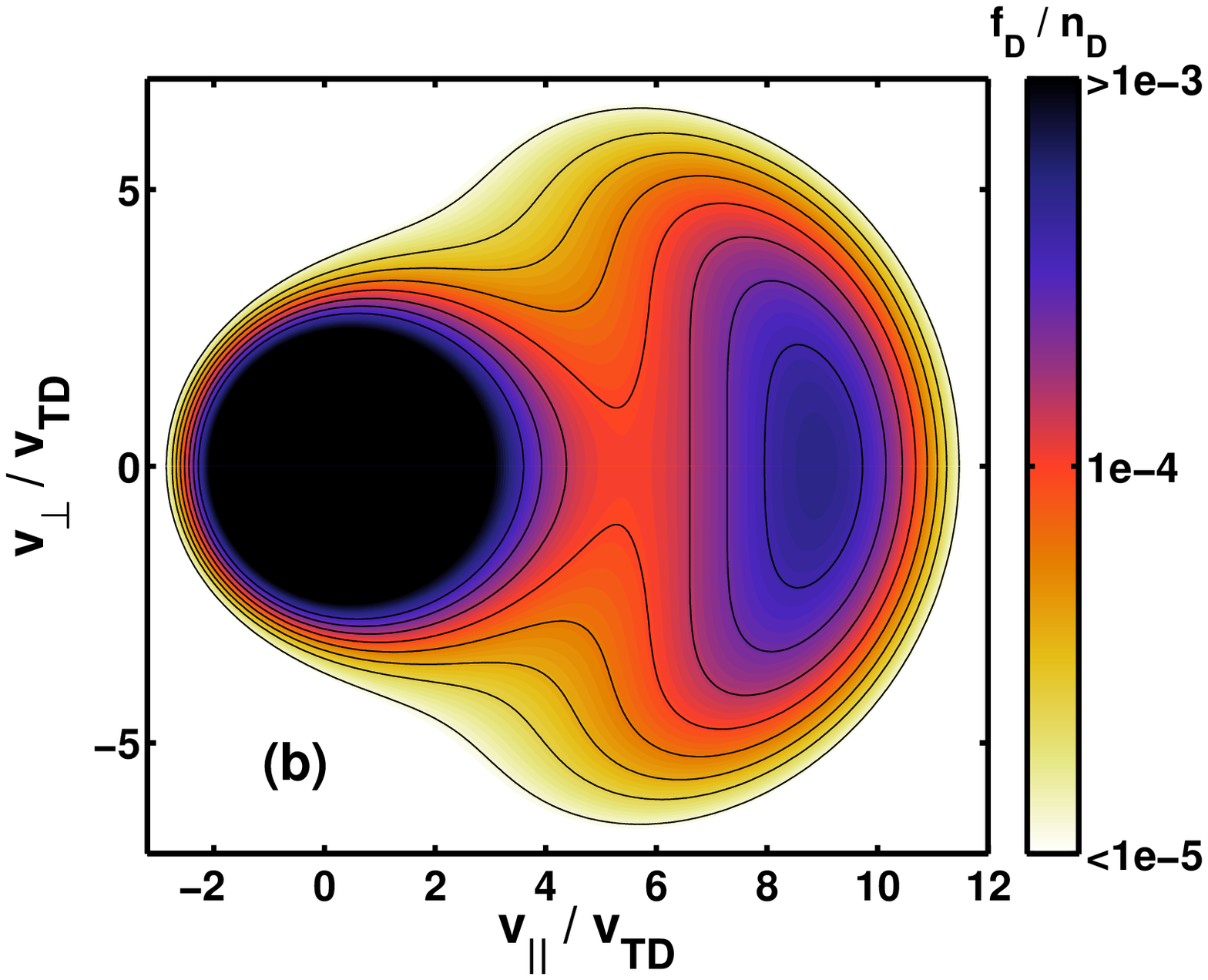}
\caption{\label{figcodiondemo} Deuterium distribution function in a
  hot plasma characterized by $T=1\,$keV for all
  particle species, \mbox{$n_e = 3\cdot10^{19}\,$m$^{-3}$} and effective
  charge $Z\sub{eff}=2$ due to fully ionized carbon impurities with
  $n_C/n_D = 4\%$. The electric field $E = 1.64\,$V/m corresponds to
  $E/E_D = 0.13$. (a) Time evolution of the $\xi = 1$ cut of the
  distribution and (b) contour plot of the steady state distribution,
  established after 20\,s.  }
\end{center}
\end{figure}

Figure~\ref{figcodiondemo}(a) shows a typical example of the evolution of
the ion distribution for a case where the electric field is above the
minimum required for runaway acceleration. The plasma parameters used
are characteristic for tokamaks with a hot bulk deuterium plasma at
$1\,$keV and fully ionized native carbon impurities. Note that the loop
voltage is typically $\lesssim 1$~V in normal tokamak operation,
corresponding to $E \lesssim 0.2$~V/m~\cite{hollmannetal2013,helanderetal2002}.  A
contour plot of the distribution in velocity space when steady state
is reached is shown in Fig.~\ref{figcodiondemo}(b).
For the parameters used here, approximately $5\%$ of the ion population 
has been accelerated and the linearization used to derive
Eq.~(\ref{eqgenkeforevolution}) is well satisfied.

The steady state distribution is typically established in 10-20\,s at this 
temperature and density, and the time-scale varies with
plasma parameters as the collision time defined in connection with
Eq.~(\ref{eqioncolop}), $\tau_{ie} \propto T_i^{3/2}/n_e$.
For stronger electric fields, the initial evolution of the
distribution can be followed, but the linearization breaks down before
the steady state can be reached.  The entire ion distribution will
eventually run away when $E\gtrsim 0.2E_D$ for the $Z\sub{eff} = 2$
case considered here, and the linearization breaks down within $\sim
30\,$ms with such an electric field.

\section{Results}
\label{seccolops}

\begin{figure}[t]
\begin{center}
\includegraphics[width=0.49\textwidth]{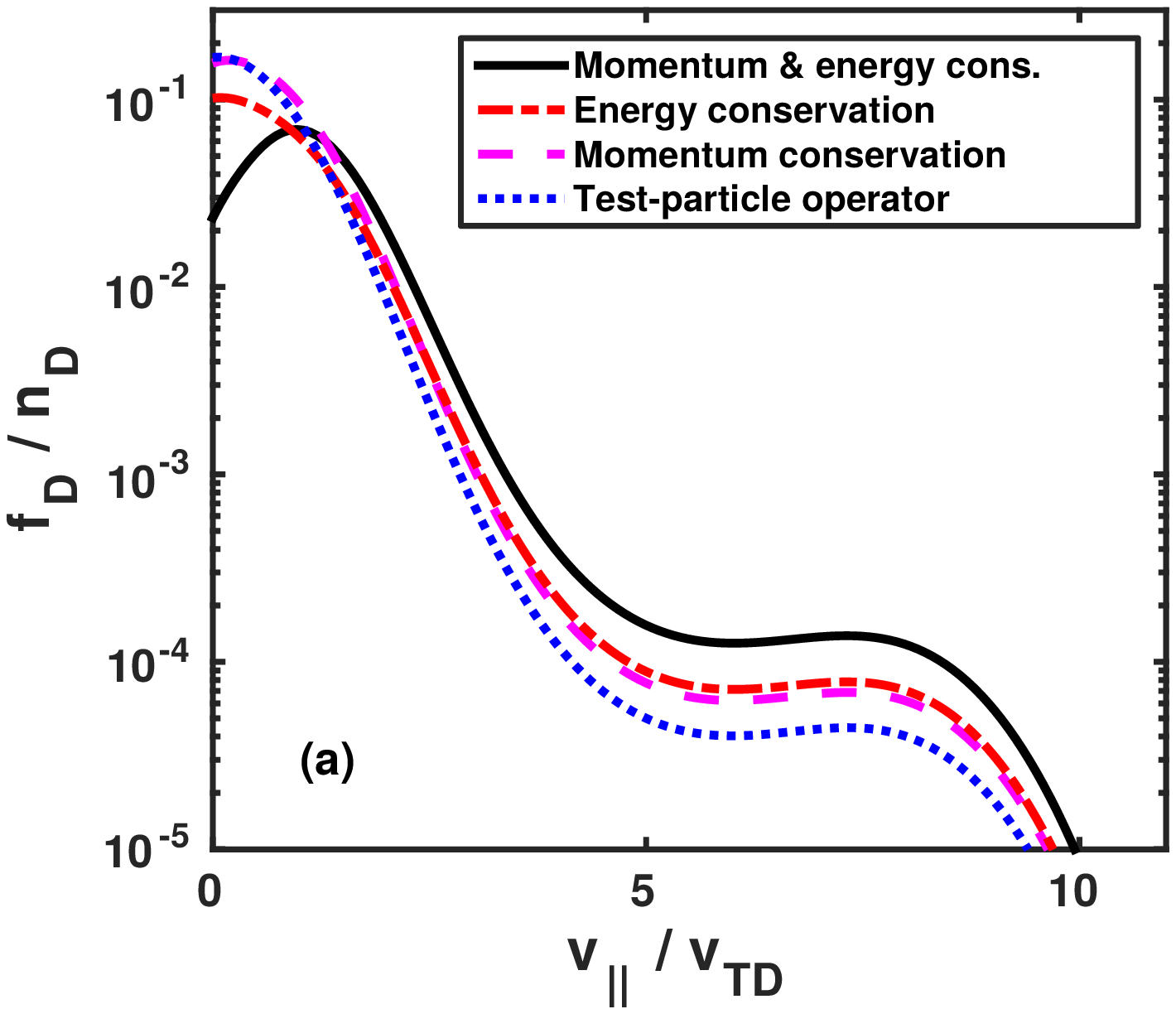}
\includegraphics[width=0.49\textwidth]{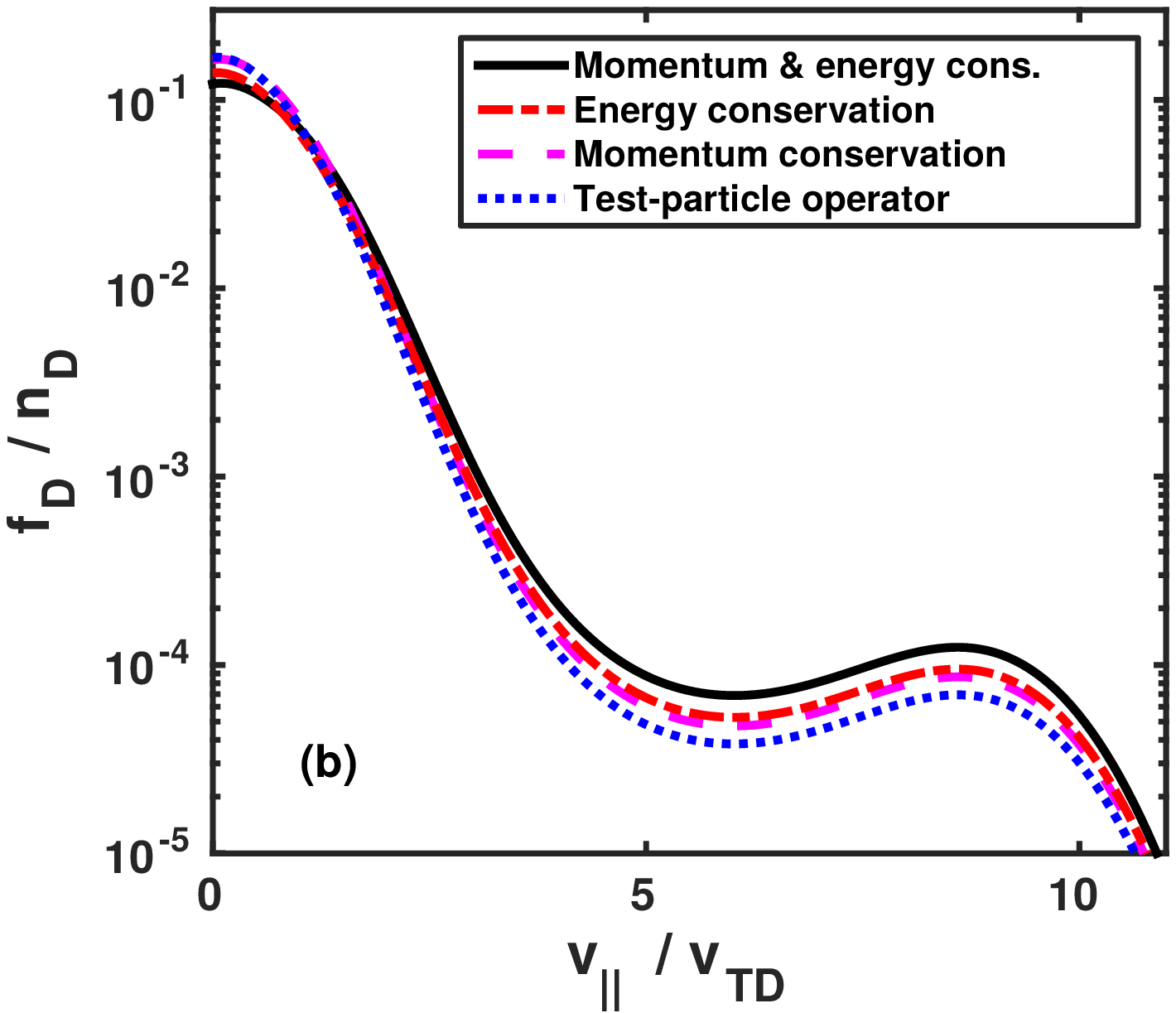}
\caption{Comparison of the $\xi=1$ cut of deuterium distributions
  calculated using \textsc{CODION}, retaining all (solid) or
  individual (dashed, dash-dotted, dotted) conservative terms in the
  ion self-collision operator. Here, $T =1$ keV for
  all species, and $n_{D^+} = 3 \cdot 10^{19}\, {\rm m}^{-3}$. The only 
  impurity is fully ionized carbon, of density such that the specified 
  effective charge is obtained, and electric fields and times are chosen 
  to produce a significant runaway population. (a) $Z\sub{eff} = 1.5$, $E
  = 2.5\,$V/m at $t=0.9\,$s, and (b) $Z_{\rm eff} = 2$, $E =
  2\,$V/m at $t= 0.8\,$s.}
\label{figcolops}
\end{center}
\end{figure}

Expansions of the collision operators appearing in
Eq.~(\ref{eqgenkeforevolution}) have been used previously in the
literature to consider ion runaway analytically.
Here we compare the effect of various approximations to the collision
operator on the numerical solution of the distribution function.
The plasma parameters of Fig.~\ref{figcodiondemo} are used as a basis
in the comparisons, but the behavior illustrated is characteristic of
a wide range of parameters.  We first consider the effect of
neglecting the conservative field-particle terms in the self-collision
operator. Figure~\ref{figcolops} shows the $\xi = 1$ cut of the
distribution of the bulk deuterium population for two
  cases, with effective charge $Z\sub{eff}=1.5$ and $Z\sub{eff}=2$,
  respectively. It can be seen that, using only the test-particle operator, the dominant
behavior of the fast-ion population as given by the fully conserving case is 
reproduced, with the main difference being the rate at
which it builds up. This is expected, since the conserving
field-particle terms are proportional to $f_{Mi} \propto
\exp(-x_i^2)$, and therefore act mainly on the thermal bulk of the
distribution.

Figure~\ref{figcolops}(a) shows a case with $Z\sub{eff}=1.5$.  With a
lower amount of impurities to which momentum and energy can be
transferred, the fully conserving linearized collision operator for
self-collisions will exhibit unphysical behavior before a significant
runaway population has time to form, which is clearly illustrated by
the distortion near $v=0$. The reason is that when the
  conserving terms are kept in the kinetic equation, the
  distribution is heated by the electric field, causing the linearized
  equation to break down after some time. This is also observed
  in the solution obtained using only the energy conserving term, albeit less pronounced. The
distribution functions obtained with the momentum conserving
self-collision operator tends to stay regular for longer.
Figure~\ref{figcolops}(b) shows a similar case but with higher
impurity content, $Z\sub{eff}=2$, for which all operators yield
well-behaving results. The main difference between the conserving
operators and only using the test-particle operator is that the former
typically lead to a runaway rate which is at least twice as large.
An additional consequence of the low effective charge was 
demonstrated in Fig.~\ref{figtestedemo}(a), where impurity ions were
shown to be more easily accelerated by an electric field than
the bulk species, implying that for low $Z\sub{eff}$ the assumption of
stationary impurity ions may be violated.

In conclusion, the high-energy part of the ion
distribution obtained using only the test-particle operator is in
qualitative agreement with the result obtained with conservative
operators, but the runaway rate is expected to be lower in the
test-particle case. A quantitative investigation of runaway rates
 for impurity species is presented in
 Section~\ref{applications}.  

\begin{figure}[t]
\begin{center}
\includegraphics[width=0.49\textwidth]{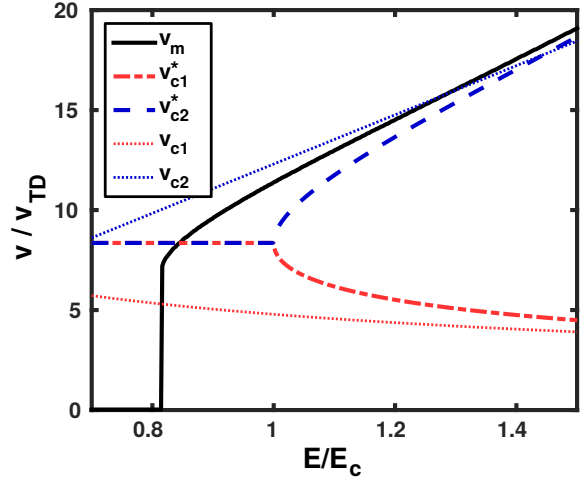}
\caption{Electric field dependence of the location $v_m$ of the
  maximum in the runaway tail, obtained using \textsc{CODION}, for a
  fully ionized impure deuterium plasma with $n\sub{C}/n\sub{D} =
  4\%$, $Z\sub{eff} = 2.5$ and equal temperature for all particle
  species. The boundary velocities $v_{c1}^*$ and $v_{c2}^*$ of the
  acceleration region obtained numerically from
  Eq.~(\ref{eqtestfriction}) (using the velocity which minimizes the 
  friction when $E<E_{ci}$), and their approximate values $v_{c1}$
  and $v_{c2}$ given by Eqs.~(\ref{eqcritvs1}-\ref{eqcritvs2}), are
  also shown. The quantities shown are independent of electron density
  and temperature. For electric fields $E\lesssim 0.8 E_{ci}$, no
  maximum forms in the runaway ion distribution. }
\label{figcomptestv}
\end{center}
\end{figure}

It is instructive to compare the behavior of the numerical solution
to the characteristic behavior indicated by the test particle
friction given in Eq.~(\ref{eqtestfriction}).
Noting that the runaway ion velocity satisfies $v_{Ti} \ll v \ll
v_{Te}$, we can expand the contributions to Eq.~(\ref{eqtestfriction})
using the known low and high-velocity limits of the Chandrasekhar function. The test
particle friction in this limit reduces to
\be
F_i^\text{test} \approx -\frac{m_i v_{Ti}}{\tau_{ie}}\left[\frac{Z_{\rm eff} + \overline{n}}{x_i} + \frac{4}{3\sqrt{\pi}}\left(\frac{T_i}{T_e}\right)^{3/2}\sqrt{\frac{m_e}{m_i}}x_i\right],
\label{eqapproxtestfriction}
\ee
where $\overline{n} = \sum_j n_j Z_j^2 m_i/n_e m_j$ allows for
arbitrary impurity content. Consider first the minimum value of the magnitude of the collisional
friction force; this will determine the minimum electric field which
can accelerate a fast test ion. Minimizing
Eq.~(\ref{eqapproxtestfriction}) yields
\begin{align}
v\sub{min} &=  v_{Te}\left[\frac{3\sqrt{\pi}}{2}\frac{m_e}{m_i}(Z\sub{eff}+\bar{n})\right]^{1/3}, \\
 F_i^\text{test}(v\sub{min}) &= -2\frac{m_i v_{Ti}}{\tau_{ie}} \frac{T_i}{T_e} \left[\frac{3}{2\pi}\frac{m_e}{m_i}(Z\sub{eff}+\bar{n})\right]^{1/3}.
\end{align}
From this it follows that the minimum, ''critical'', value $E_{ci}$ of the electric
field above which a test ion can be accelerated is given by
\be
\frac{E_{ci}}{E_D} = \frac{Z_i\left(Z_{\rm eff} + \overline{n}\right)^{1/3}}{\left| 1 - Z_i/Z_{\rm eff}\right|}\left(\frac{3}{2\pi}\frac{m_e}{m_i}\right)^{1/3}.
\label{eqecrit}
\ee
By taking $Z_i e E^* + F_i^\text{test} = 0$, we can find the range of
test ion velocities, $v_{c1} < v < v_{c2}$, for which acceleration in
a given electric field occurs, as discussed in
Refs.~\cite{holman1995,furthrutherford1972}. Using the expression for
the friction given in Eq.~(\ref{eqapproxtestfriction}) results in a
third order equation, however simpler approximate formulae can be
obtained by noting that $v_{c1}$ will fall near to the region
dominated by ion friction, and $v_{c2}$ in the region dominated by
electron friction.  Retaining only the corresponding terms in
Eq.~(\ref{eqapproxtestfriction}) yields, for arbitrary impurity
content,
\bea
v_{c1} &=& v_{Ti}\,\sqrt{\frac{Z_i T_e}{2T_i}\left(\frac{E}{E_D}\right)^{\hspace{-1mm}-1}\hspace{-2mm}\frac{Z_{\rm eff} + \overline{n}}{\left|1-Z_i/Z_{\rm eff}\right|}}\,, \label{eqcritvs1}\\
v_{c2} &=& v_{Te}\frac{3\sqrt{\pi}}{2}\frac{E}{E_D}\frac{\left| 1 - Z_i / Z_{\rm eff}\right|}{Z_i} .
\label{eqcritvs2}
\eea
These equations generalize the corresponding expressions in
Ref.~\cite{holman1995} to arbitrary plasma composition.  Note that
these formulae are only valid when $E$ is sufficiently large compared
to $E_{ci}$, since at $E=E_{ci}$ ion and electron friction contribute
equally.  We may expect that in steady-state, the position of the
high-velocity maximum of the distribution function, denoted $v_m$, is
close to $v_{c2}$, which scales linearly with $E$ in the approximate
form given by Eq.~(\ref{eqcritvs2}).
This is confirmed numerically and illustrated in
Fig.~\ref{figcomptestv}, where we show the variation with electric field of $v_m$,
obtained from steady-state \textsc{CODION} solutions of
Eq.~(\ref{eqgenkeforevolution}).  Also shown are the boundary
velocities of the acceleration region, resulting from numerical
solution of the force balance using the full test particle friction,
Eq.~(\ref{eqtestfriction}), as well as their approximate forms
Eqs.~(\ref{eqcritvs1}-\ref{eqcritvs2}). The values converge when the
system is strongly driven by a large $E$.
The linear dependence of $v_m$ is clearly visible at large $E$, where
it approaches the value given by the test particle approximation.
The analytic approximation for $E_{ci}$, Eq.~(\ref{eqecrit}),
is only accurate to $\sim 20\%$, however, indicating that collisional diffusion
contributes significantly to the evolution at lower electric fields.
Since the linearization breaks down more rapidly at larger electric
fields, it is mainly at fields near the threshold for runaway
generation that the model can consistently be applied to study the
long-term evolution of the ion runaway tail, making a full kinetic
simulation essential for capturing the important physics.  For the
more massive impurities, the features of the test-particle model
become increasingly accurate since the runaway ion population is
further separated in velocity space from the thermal bulk.

It is important to point out that neither the diffusion terms nor the
field-particle self-collisions have been accounted for in the
derivations of the above estimates, which are meant to give simple
expressions that show how the essential quantities scale with the plasma
parameters, and to provide a useful physical picture for illustrating
the ion runaway phenomenon. A complete description will be provided
only by numerical solution of the kinetic equation.


\section{Applications}
\label{applications}
In this section, \textsc{CODION} is applied to calculate runaway ion
distributions for typical solar flare and fusion plasmas. The rate at
which a fast ion population forms due to the runaway mechanism is
determined, and it is investigated whether the difference in
acceleration rate between different ion species can explain the
enhanced abundance of heavy ions in the solar wind. We also
consider the possibility of Alfv\'enic instabilities being driven by
runaway ions during tokamak disruptions.

Throughout this section, time-scales are chosen so that significant
fast ion populations have time to form, which typically takes a few hundred
ion-electron collision times. We define the runaway density, $n_r =
(1/n_i)\int_{v>v_{c1}^*} \hspace{-1mm} \rd^3 v \, f_i$, as the fraction
of ions with velocity larger than the low-velocity numerical solution
of $Z_i e E^* + F_i^\text{test} = 0$, denoted $v_{c1}^*$, which if
$E<E_{ci}$ is taken as the velocity $v\sub{min}$ minimizing the
friction $F_i^{\text{test}}$.


\subsection{Ion acceleration in solar flares}
Solar flares are thought to be initiated by reconnection in the
corona~\cite{liuwang2009}, but the origin of the observed fast ion
populations in flares is still not completely
understood~\cite{garrardstone1994}.
Both stochastic acceleration by waves and the direct acceleration of
the particles by the electric field have been considered, and it
appears likely that a combination of the two can be at
work~\cite{holman1995,braddyetal1973,sylwesteretal2014}.

The effective accelerating field experienced by a given species varies
with its charge and the effective charge of the plasma, as discussed
in Section~\ref{secridistrib}.  This can give rise to preferential
acceleration of heavier elements under certain circumstances, and this effect
was considered in~\cite{holman1995}, where estimates of the runaway
rate were given based on the approximate formula
of~\cite{gurevich1961}.  With \textsc{CODION} we can determine
the time evolution of the ion distribution function numerically, and
evaluate the dependence of the acceleration rate on various ion
parameters.

The composition of the solar plasma, particularly the metallic
elements, has been studied extensively in recent years, however much
uncertainty remains.  We will choose parameters consistent with the
choices made by Holman \cite{holman1995}. We use the plasma temperature
$T = 700\,$eV for all particle species, and hydrogen density $n\sub{H} =
3\cdot10^{17}\,$m$^{-3}$. Elements with atomic number $Z\leq 6$ can
readily be assumed to be fully ionized at this temperature.  The
plasma composition is based on the ion abundance recommended by
Schmeltz \emph{et al.} \cite{schmelzetal2012}. We use a helium
population of density $n\sub{He}/n\sub{H} = 6\%$, and represent all
heavier impurities by a carbon population of density
$n\sub{C}/n\sub{H} = 0.1\%$, corresponding to an effective charge
$Z\sub{eff} = 1.13$. Electric field strengths in solar flares are not
well constrained by experimental observation, and we will investigate
the rate of acceleration at a range of values. The Dreicer field is
$E_D = 224\,$mV/m for this set of plasma parameters.

\begin{figure}[t]
\vspace{-3mm}
\begin{center}
\includegraphics[width=.9\textwidth]{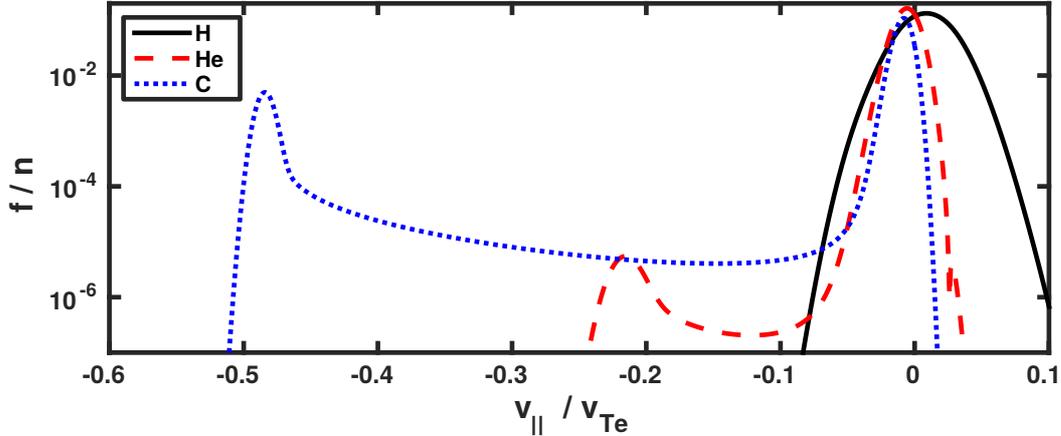}
\caption{\label{multirunaway flare} Distribution functions of
  hydrogen, helium and carbon after 30\,s of acceleration in a solar
  flare-like plasma, with $E=50\,$mV/m. All distribution functions are
  normalized to their respective densities. The temperature is $T=700\,$eV for all species, 
  $n\sub{H} = 3\cdot10^{17}\,$m$^{-3}$, $n\sub{He}/n\sub{H} = 6\%$, and 
  all other impurities are represented by a carbon population with $n\sub{C}/n\sub{H} = 0.1\%$, yielding
  $Z\sub{eff} = 1.13$. The runaway densities are
  $n_{r,\text{H}} \approx 0$, $n_{r,\text{He}}=3.7\cdot 10^{-4}$ and
  $n_{r,\text{C}}=0.18$. \vspace{-.1cm} }
\end{center}
\end{figure}

The critical electric fields $E_{ci}$ for the ion species in such a
plasma are $E_{c,\text{H}} = 154$\,mV/m for hydrogen,
$E_{c,\text{He}}= 40$\,mV/m for helium and $E_{c,\text{C}}= 20\,$mV/m
for carbon.  Note that the acceleration rate depends strongly not only on
$E/E_{ci}$, but also on $v_{c1}/v_{Ti}$. Figure \ref{multirunaway
  flare} shows the $v_\perp = 0$ cut of the distribution functions of
hydrogen ($^1$H), helium ($^4$He) and carbon ($^{12}$C) after 30\,s of acceleration from initial
Maxwellians, with the plasma parameters specified above and an
electric field $E=50\,$mV/m. This is significantly below the hydrogen
critical field, and no hydrogen runaway population forms.  Runaway ion
populations of both helium and carbon do form however, with runaway
densities $n_{r,\text{He}} = 0.037\%$ and $n_{r,\text{C}} = 18\%$, respectively.

\begin{figure}
 \includegraphics[width=0.9\textwidth]{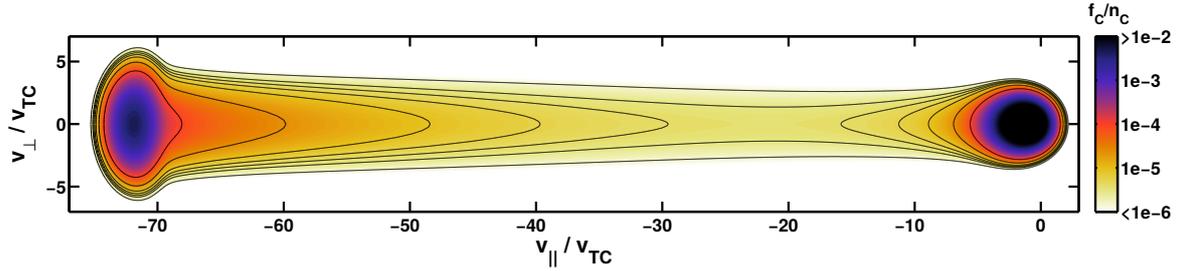}
\caption{\label{flare carbon contour} Distribution function $f_C$ of
  carbon, accelerated from an initial Maxwellian for $30\,$s by an
  electric field $E=50\,$mV/m, in a fully ionized solar flare-type
  plasma with $n\sub{H} = 3\cdot10^{17}$m$^{-3}$, $n\sub{He}/n\sub{H}
  = 6\%$ and $n\sub{C}/n\sub{H} = 0.1\%$ , and temperature $T=700\,$eV
  for all species. }
\end{figure}

Positive values of $v_\parallel$ represent the direction of the
electric field. Therefore, Fig.~\ref{multirunaway flare} also
illustrates how heavier ions (charge $Z>Z\sub{eff}$) are accelerated
in the direction opposite to the electric field, dragged by electron
friction. The corresponding 2D carbon distribution function is shown
in Fig.~\ref{flare carbon contour}, displaying a strong directional
anisotropy (compare to Fig.~\ref{figcodiondemo}(b)). This can be
understood by the observation that the accumulation velocity $v_{c2}$
is located at a higher value of $x_i=v/v_{Ti}$ for heavier ions. Since
pitch angle scattering of the energetic heavy ions scales with
velocity like $Z\sub{eff}/x_i^3$, the mechanism will be less
effective than for light bulk ion species in increasing the
perpendicular energy of the distribution.

We will now investigate how the rate of acceleration varies with
electric field strength for different ion species. To determine the
runaway density of heavier ions, we have introduced trace amounts of
each ion species with charge between 2 (helium) and 18 (argon),
assumed to be fully ionized. In practice, ions of charge $Z_i>8$ will
typically not be fully ionized at the temperature considered due to their high ionization energy,
meaning that the results shown here will overestimate the acceleration
rate of the heavier ions. The ion masses have been set to that of the
most common isotope, i.e. $^7$Li, $^9$Be, $^{20}$Ne etc. Both $^3$He
and $^4$He are shown, with $^3$He showing a significantly enhanced
runaway rate as compared to that of $^4$He, for all values of $E$.

\begin{figure}[t]
\begin{center}
\includegraphics[width=.49\textwidth]{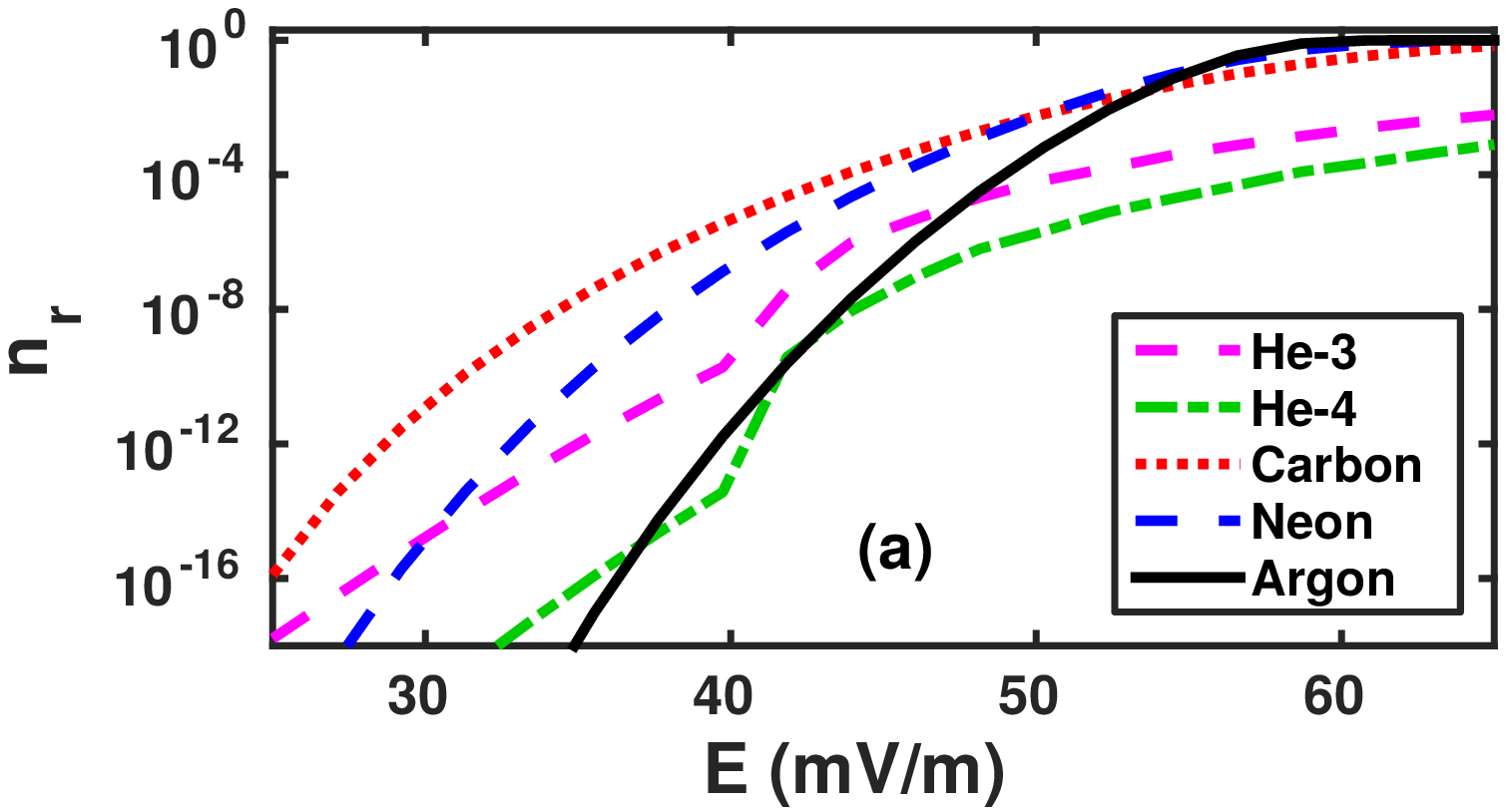}
\includegraphics[width=.49\textwidth]{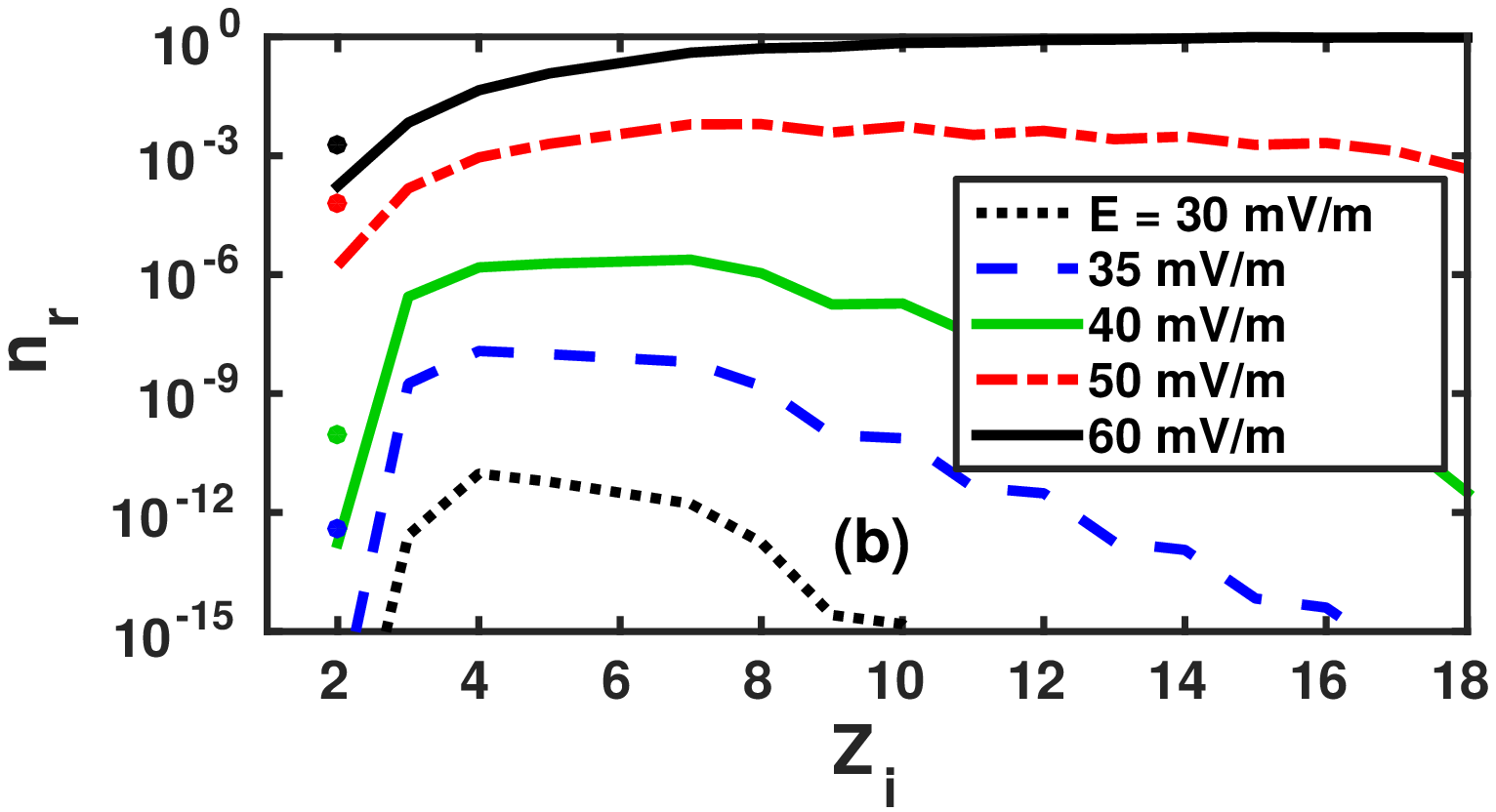}
\caption{\label{flare runaway rate} Solar flare-like plasma with all
  particle species at the same temperature $T=700\,$eV. The
  composition is mainly hydrogen, $n\sub{H} =
  3\cdot10^{17}\,$m$^{-3}$, and helium-4, $n\sub{He}/n\sub{H}=6\%$,
  with carbon of density $n\sub{C}/n\sub{H} = 0.1\%$ representing all
  other impurities, each appearing in trace amounts. Ions are
  accelerated for 1\,s from initial Maxwellians. Due to the low
  effective charge $Z\sub{eff} = 1.13$, the hydrogen background
  remains stationary ($E^*$ is small), and heavier elements are assumed
  to be present only in trace amounts, making self-collisions
  negligible. a) Runaway density $n_r$ as a function of electric field
  strength $E$ for various ion species.  b) Runaway density as
  a function of ion charge $Z_i$. For each atomic number the most common
  isotope is used, except for helium ($Z_i=2$) where both $^3$He and
  $^4$He are shown, with the $^3$He runaway density indicated by
  dots.}
\end{center}
\end{figure}

Figure \ref{flare runaway rate}(a) shows how the runaway density $n_r$
depends on electric field after 1\,s of acceleration for various ion
species present in the solar flare plasma. The figure illustrates how
the runaway rate is sensitive to ion parameters; at low electric
fields the heavier ions tend to be accelerated at a lower rate than
light ions, while at higher fields they are the most readily
accelerated. Note that above the critical electric field, $E_{ci}
\approx 40\,$mV/m for helium, the runaway rate increases significantly
faster than for $E<E_{ci}$. The runaway rate of $^3$He is seen to be orders of 
magnitude higher than the runaway rate of $^4$He for all electric fields considered.

Finally we illustrate the dependence of the acceleration rate on ion
charge and mass. Figure \ref{flare runaway rate}(b) shows the runaway
density $n_r$ after 1\,s of acceleration as a function of ion charge
$Z_i$ for various electric fields.  Ions of charge between $Z_i=4$
($^{9}$Be) and $Z_i=8$ ($^{16}$O) are seen to be preferentially
accelerated over lighter or heavier elements for low electric
fields. For $Z_i>8$, the trend depends on electric field. For low
electric fields, the runaway rate decreases with charge, while for
larger electric fields (in this case above approximately 50\,mV/m),
heavier ion species may be accelerated at a higher rate
than the lighter species.  Further studies are required to determine
the full effects of variation in background composition, temperature,
density, and charge states on the relative rates of acceleration
between different ion species. The results presented here demonstrate
the utility of \textsc{CODION} for the problem.

As previously noted, acceleration by quasi-static electric fields is
not the only mechanism for ion acceleration in a flare
plasma. Interaction with Alfv\'en waves can accelerate ions which have
velocities above the Alfv\'en velocity.  As this is usually well above
the thermal ion velocity, an initial acceleration by electric fields
may be required before the process becomes significant
\cite{harrison1960,holman1995}. \textsc{CODION} provides the means for
more accurate modeling of the effects of such interactions.


\subsection{Tokamak disruptions}

During tokamak disruptions the plasma temperature drops from the
typical operating regime of around several keV to a few eV in a couple of
milliseconds.  A large electric field is initially induced parallel to
the magnetic field to maintain the plasma current of several MA, potentially
leading to the formation of a beam of energetic electrons through the runaway mechanism. 
The potential for damage by such a focused high-energy beam on 
contact with the vessel wall is large, and runaway generation must as 
far as possible be suppressed.  To study the physics and mitigation of 
runaway electrons, disruptions can be induced by the injection of
large quantities of noble gas, often in amounts comparable to the initial plasma
inventory or larger~\cite{hollmannetal2013}.

The large induced electric field will usually decay rapidly on a
timescale of a few ms in response to the formation of a narrow runaway
electron (RE) beam.  With runaway electrons reaching energies of order
tens of MeV, they can carry a significant fraction of the
pre-disruption plasma current and can drive high frequency
electromagnetic instabilities through resonant
interactions~\cite{fulopetal2006,pokoletal2008,komaretal2013,pokoletal2014}.
Recently, low-frequency magnetic fluctuations in the range $f \approx
60-260\,$kHz, have been observed in the TEXTOR tokamak during induced-disruption 
studies with argon massive gas injection (MGI). These
fluctuations take the form of either a strong signal at a distinct
frequency~\cite{pappeps2014}, or accompanied by broadband
activity~\cite{zengetal2013}.  The fluctuations appear to limit the RE
beam formation in these cases, as the magnetic perturbations may
scatter the runaway electrons and provide passive mitigation. Aside
from the potential consequences for mitigation, observed instabilities
offer a non-intrusive diagnostic for both bulk plasma and fast-particle 
properties, through the extensively applied technique of MHD
spectroscopy~\cite{holtiesetal1997}.

Fast ions resulting, for example, from certain heating schemes are
well known to resonantly drive low frequency Alfv\'{e}nic
instabilities in typical operational
scenarios~\cite{heidbrink2008,zoncaetal2007}. Runaway ions may thus
also provide a potential drive for the fluctuations observed.
Interestingly, toroidal Alfv\'{e}n eigenmodes
(TAEs)~\cite{chengchance1986} can have frequencies and mode numbers in
the same range as the post-disruption magnetic activity. Therefore,
the excitation of TAEs by runaway ions was recently considered in
Ref.~\cite{fulopnewton2014} using an analytical approximation for the
runaway distribution function. As the validity of the approximate
distribution was limited, definite conclusions could not be drawn.
With \textsc{CODION} we can extend the study using the numerically
calculated ion distribution. The cold post-disruption plasma is highly
collisional, motivating the use of a homogeneous background plasma and
the neglect of magnetic trapping when evaluating the effective
electric field.

The TAE perturbation is typically dominated by two neighboring
toroidally coupled harmonics at large aspect ratio, with poloidal mode
numbers $m$ and $m+1$.  The mode is localized about the minor radius
$r=r_0$ at which the magnetic safety factor is $q_0=(2m+1)/2n$, where
$R_0$ is the radius of the magnetic axis and $n$ is the toroidal mode
number.  The TAE frequency is $\omega = v_A/(2 q_0 R_0)$, where
$v_A=B/\sqrt{\mu_0 \rho_m}$ is the Alfv\'{e}n speed and $\rho_m$
the mass density.  The two component harmonics allow resonant
interaction with particles whose parallel velocity $v_\parallel$
satisfies $|v_\parallel|\simeq v_A/3$ or $|v_\parallel|\simeq v_A$.
It was argued in Ref.~\cite{fulopnewton2014} that as the runaway ions
accelerate, the inverted region of their energy distribution,
$\partial f/\partial \mathcal{E} > 0$ where $\mathcal E=m_{{i}} v^2/2$
is the particle energy, can reach the lower Alfv\'{e}n resonance,
$v_\parallel=v_A/3$ and may drive the TAE.  If the radial runaway ion
profile peaks on axis, the spatial gradient $\partial f/\partial r$
will give an additional positive contribution to the growth rate.
Taking parameters characteristic of argon MGI-induced disruptions,
$n\sub{D} = 3\cdot10^{19}\,$m$^{-3}$, $n\sub{Ar} = 0.1 n\sub{D}$,
$Z\sub{Ar} = 2$, and anticipating a native background carbon impurity
with $n\sub{C}=0.08 n\sub{D}$ and $Z\sub{C} = 2$ so that
$Z_{\text{eff}} = 1.26$, a background ion temperature of $T_i=10\;\rm
eV$, toroidal magnetic field $B= 2\,$T and major radius $R_0 =
1.75\,$m means that the resonance condition $v_\parallel=v_A/3$
requires deuterium ions with velocities $v \simeq 35 v_{T\text{D}}$.

The electric field required to accelerate bulk ions to the resonant
velocity at these low temperatures is substantial, varying in response
to changes in $Z_{\rm eff}$ but is typically {$\gtrsim 0.3 E_D \sim
  100\,$V/m}.  Such field strengths are unlikely to occur during a
disruption, and they would be short-lived if they did \cite{smithhelander2006}.
Therefore, we conclude that whilst ion runaway may be of interest in
hot fusion plasmas, runaway ions are unlikely to provide the drive for
the observed fluctuations during disruptions.

\begin{figure}[h]
\begin{center}
\includegraphics[width=.55\textwidth]{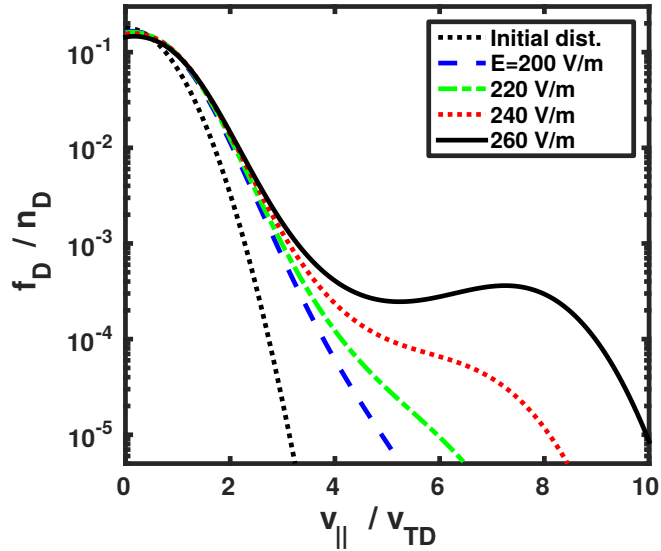}
\caption{\label{cold TEXTOR E vary} Deuterium distribution function
  resulting from various electric fields after $t=2\,$ms of
  acceleration from an initial Maxwellian in a TEXTOR
  MGI post-disruption-like plasma, with $n\sub{D} = 3\cdot10^{19}\,$m$^{-3}$,
  $n\sub{C} = 0.08n\sub{D}$, $n\sub{Ar} = 0.1n_D$ and $T=10\,$eV for all
  species, with $Z\sub{C} = Z\sub{Ar} = 2$. The critical field for deuterium is $E_{cD} = 295\,$V/m, and the Dreicer field is $E_D = 962\,$V/m. }
\end{center}
\end{figure}

To quantify the electric field needed for significant ion runaway, we
show in Fig.~\ref{cold TEXTOR E vary} how the deuterium distribution
evaluated after $2\,$ms of acceleration from an initial Maxwellian --
a typical time scale for the induced electric field -- varies with (a
constant) electric field. The parameters are $n\sub{D}=3\cdot
10^{19}\,$m$^{-3}$, $T=10$\,eV and the same plasma composition as before with
$Z\sub{eff}=1.26$. It is seen that for electric fields below $\sim
200$\,V/m, no runaway tail tends to form, and even with $E=260$\,V/m the
fast ions are far from the resonant velocity near $35 v_{T\text{D}}$. The
behavior is sensitive to which temperature is chosen for the plasma:
increased temperature decreases the electric fields needed to
accelerate ions, but makes the acceleration timescale longer.

Note that a higher amount of assimilated argon, or a weaker magnetic
field, would lead to a lower Alfv\'en velocity and TAE frequency, and
therefore the runaway ions would reach the resonance condition more
easily. The higher electron density leads to an increased
collision frequency, however, so that the runaway ion distribution 
requires a longer time to form. The level of argon absorption during mitigation
varies between machines~\cite{hollmannetal2013} and the absorption
appears to decrease for bigger machines. Therefore, although we may
expect a higher electric field in large tokamaks such as ITER, the
higher bulk plasma density and likely lower absorption suggests no ion
runaway will occur.  Note that as the ITER magnetic field is so high,
even if the assimilated argon was equal to the initial deuterium
inventory, the ions would need to reach $\sim 40 v_{Ti}$ (at 10 eV) to
reach the typical $v_A/3$ resonance, assuming $B=5\,$T, $R_0 = 6\,$m
and $n\sub{D} = 1\cdot 10^{20}\,$m$^{-3}$.

Finally we note that the model presented in this paper assumes the
initial ion distribution to be a stationary Maxwellian.  Fast ion
populations present due to heating schemes in use before the
disruption may not be completely expelled, and it is not certain that
the initial distribution is accurately described by a
Maxwellian. The dynamics of ion acceleration starting from these
non-Maxwellian distributions may yield a fast-ion population more
readily.  Furthermore, since plasma conditions change on a time scale
of a few collision times during the sudden cooling associated with
instabilities or disruptions, so-called ``hot-tail'' runaway
generation may occur. This has been shown to be an important effect
for runaway of electrons, where a seed of runaway electrons are provided
by fast electrons present before cooling \cite{cooling}. These fast
electrons are cooled at a slower rate than the low-energy electrons,
and may find themselves in the runaway region when the plasma has
reached its final temperature. However, using \textsc{CODION} to
investigate the effect of hot-tail ions, it has been concluded that
the effect is small for realistic fields. The reason is that, if the
electric field is not high enough for the ions to overcome the
friction and become runaways in the first place, all hot-tail
ions will necessarily also be slowed down. A low-velocity inverted ion
population may however form as a result of the cooling process even
for such low electric fields, with its peak near the velocity which
minimizes the collisional friction at the final temperature (typically
around 6-8 thermal velocities). This is still significantly lower than that
needed for resonant interaction with Alfv\'en waves.

One refinement to the model would be the inclusion of ``knock-ons'',
i.e. large-angle collisions, which have been neglected in the Fokker-Planck
equation. It is well known that single collisions can 
change the momenta of the interacting particles significantly, and a runaway ion
interacting with a bulk ion could cause both to end up in the runaway
region. In a situation where the electric
field is low enough that runaway ions are produced at a low rate
through the standard acceleration mechanism, knock-on collisions could
possibly contribute significantly to the runaway generation rate. This
has been demonstrated to be the case for electron runaway, where this
effect drastically affects the rate at which runaways are produced. A
simplified runaway ion knock-on operator could potentially be
constructed from the Boltzmann collision integral under the assumption
that fast ions accumulate near $v_{c2}$ and collide mainly with the
bulk distribution, since the fast ion distribution is assumed to be a
small perturbation in our linearized model.  However, there are
differences between ion runaway and electron runaway that suggest that
knock-on runaway generation is a less significant effect for ions
than for electrons. Since our linearized model restricts the study to
cases where $E\sim E_{ci}$ (which is also the regime where knock-on 
generation would be expected to be significant), the accumulation velocity
near $v_{c2}$ will not be significantly larger than the runaway
velocity $v_{c1}$. Because of this, collision events where both particles
end up in the runaway region will be less frequent. This is in contrast 
to electron runaway, where the electrons have unbounded energy (neglecting radiation
effects).

There are applications for knock-on operators other than avalanche
generation. It has been suggested that fast ion
populations due to other sources -- for example hot alpha particles
created in fusion reactions or ions heated by external sources such
as neutral beam injection (NBI) or radio frequency (RF)-heating --
could accelerate bulk impurity ions, which could in turn be used for
diagnostics \cite{knock-on alphas, knock-on helander}. The suggested
collision operator could be implemented in CODION, and the
time-evolution of bulk impurities solved for in the vicinity of an
assumed or numerically obtained background of fast ions, however
this is outside the scope of the present paper.

%


\section{Conclusions}
\label{conclusions}

Electron runaway resulting from the occurrence of a strong electric
field in a plasma has been the subject of extensive study, and
numerical tools exist to simulate the electron dynamics.
The analytic description of the associated ion acceleration was
developed at the same time, but its application has been much more
limited and is restricted by the various approximations to the
collision operator which were required.

We have developed an efficient open-source numerical tool,
\textsc{CODION} \cite{CODION github}, which solves the ion Fokker-Planck 
equation as an initial value problem in a fully ionized plasma.
A uniform background magnetic field is assumed, along with initially
stationary Maxwellian distributions, however arbitrary impurity densities and
temperatures may be specified.  A model operator for ion self-collisions based on
that used in the gyrokinetic code GS2 \cite{GS2 paper} has been
employed, satisfying momentum and energy conservation, non-negative
entropy production and self-adjointness.
A simplified analytical model based on the large mass ratio is used
for ion-electron collisions, allowing a description of ion-electron
friction caused by the perturbation of the electron distribution due
to the electric field.
However, we wish to note that our model will break down for strong
electric fields, as the electrons -- which are assumed to be in force
balance with the electric field and ion friction -- will be rapidly accelerated by electric fields
approaching the Dreicer field.  A full description of such scenarios
would require the simultaneous evolution of the ion and electron
distributions, for example by coupling the \textsc{CODION} and
\textsc{CODE}~\cite{codepaper} codes.

The effect of various approximations to the collision operator
commonly used in the literature has been studied numerically.
It has been demonstrated that the addition of momentum and energy
conservation in self-collisions mainly acts to increase the rate at
which the fast ion population builds up, while the qualitative
behavior is largely unaffected. For strong electric
fields, the test particle description is seen to reproduce the
characteristic velocity achieved by the runaway population well, and
the predicted critical electric field for ion acceleration is accurate
to $\sim 20\%$. Using the test particle approximation, we derived
concise analytic expressions for the critical electric field for ion
runaway, Eq.~(\ref{eqecrit}), and the typical runaway energy,
Eqs.~(\ref{eqcritvs1}-\ref{eqcritvs2}), and tested these
expressions against direct numerical simulation with CODION.

The output of \textsc{CODION} is the evolution of the 2D velocity
space ion distribution.  The utility of this has been demonstrated for
calculating acceleration rates of ions in solar flare plasmas.  The
rate at which ions are accelerated has been evaluated for a range of
ion masses and charges for a solar flare scenario based on that
considered by Holman \cite{holman1995}, and an exponential dependence
of acceleration rate with charge for $Z > 8$ has been illustrated for
this scenario.

Low-frequency instabilities, in the range characteristic of TAE modes,
have been observed in post-disruption tokamak plasmas, where the
disruption was induced by massive gas injection.  Using
\textsc{CODION} we have considered the potential for ions accelerated
in the disruption-induced electric field to drive such modes
resonantly.  The post-disruption discharge parameters are not well
constrained experimentally and simulations of a range of values
indicate that ion acceleration is possible.  The typical maximum ion
velocity achieved is too low for resonant interaction to
occur, however, and the rate of runaway generation is too slow for a
significant runaway density to be reached in the short-lived electric
fields of a typical disruption.


\section*{\normalsize {\sffamily ACKNOWLEDGEMENTS}}

\noindent The authors are grateful to Gergely Papp, Matt Landreman,
Istv\'an Pusztai and Joan Decker for fruitful discussions.  This project
has received funding from the Knut and Alice Wallenberg Foundation and
the RCUK Energy Programme [under grant EP/I501045].


\end{document}